**Free coherent evolution of a coupled atomic spin system initialized by electron scattering**

Lukas M. Veldman[1], Laëtitia Farinacci[1], Rasa Rejali[1], Rik Broekhoven[1], Jérémie Gobeil[1], David Coffey[1], Markus Ternes[2,3], Alexander F. Otte[1*]

[1] Department of Quantum Nanoscience, Kavli Institute of Nanoscience, Delft University of Technology, Lorentzweg 1, 2628 CJ Delft, The Netherlands

[2] RWTH Aachen University, Institute of Physics, D-52074 Aachen, Germany

[3] Peter-Grünberg-Institute, Forschungszentrum Jülich, D-52425 Jülich, Germany

[*] To whom correspondence should be addressed; E-mail: a.f.otte@tudelft.nl

**Full insight into the dynamics of a coupled quantum system depends on the ability to follow the effect of a local excitation in real-time. Here, we trace the coherent evolution of a pair of coupled atomic spins by means of scanning tunneling microscopy. We use a pump-probe scheme to detect the local magnetization following a current-induced excitation performed on one of the spins. Making use of magnetic interaction with the probe tip, we are able to tune the relative precession of the spins. We show that only if their Larmor frequencies match, the two spins can entangle, causing the excitation to be swapped back and forth. These results provide insight into the locality of electron-spin scattering, and set the stage for controlled migration of a quantum state through an extended spin lattice.**

One of the longstanding goals in experimental physics is the ability to create a *quantum simulator*: an engineered system of coupled quantum degrees of freedom that can be initialized in an arbitrary state, allowing one to observe its intrinsic free evolution [1]. In principle, scanning tunneling microscopy (STM) offers each of these ingredients. Individual magnetic atoms can be spatially arranged and studied by means of spin-polarized tunneling [2,3] and electron tunneling spectroscopy [4,5], allowing to respectively probe their local magnetization state and energy configuration. However, due to their slow timescales, these techniques have been able to observe the dynamic spin processes only indirectly [6,7,8].

In recent years, the STM toolbox was expanded to include pump-probe spectroscopy -- allowing spin relaxation to be probed on the nanosecond timescale [9,10] -- as well as electron spin resonance performed locally at the probe tip (ESR-STM) [11]. ESR-STM, combined with AC pulsing schemes, enabled the observation of the coherent time evolution of a single atomic spin [12], on par with achievements in semiconductor spin qubits [13,14] and NV centers [15]. However, in order to demonstrate free evolution of a pair of entangled spins, the initial excitation has to be sufficiently fast to compete with the intrinsic dynamics set by the coupling strength. ESR-STM uses a Rabi flop process for initialization, the rate of which is limited by the radio-frequency (RF) power available at the probe tip.

Here, by sequentially combining ESR-STM and pump-probe techniques, we demonstrate the detection of free, coherent flip-flop evolution of two coupled spin-1/2 atoms resulting from a nearly instantaneous electron scattering process. Using the energy resolution of ESR-STM we tune the eigenstates of two coupled spin-1/2 particles from Zeeman states $|↑↑⟩$, $|↓↑⟩$, $|↑↓⟩$, $|↓↓⟩$ to singlet-triplet states $|↑↑⟩$, $|−⟩$, $|+⟩$, $|↓↓⟩$ by varying the tip height [11,16,17]. Subsequently, using a pump-probe scheme, we excite and read out the spin projection of one of the two spins with nanosecond resolution. We observe an oscillating magnetization for the spin underneath the tip which we attribute to a flip-flop interaction between the two spins. This implies that the excitation process due to tunneling electron scattering is local: it only consists of a spin-flip on the atom underneath the tip, irrespective of the energy eigenstates of the system. This is a noteworthy result in the light of previous works, where it was deemed sufficient to consider electron-induced spin excitations as transitions between energy eigenstates [18,19,20].

We use a low-temperature STM to manipulate individual hydrogenated Ti atoms, henceforth referred to as TiH, on top of bilayer MgO islands on a Ag(100) crystal. To obtain spin polarization, we deposit Fe atoms and transfer them to the tip apex [21]. The ESR and pump-probe experiments are performed by applying the RF voltage and pulse trains to the tip, at temperatures of 1.5 K and 400 mK, respectively. We use an external field $B_{ext}$ = 450 mT in-plane at 14° angle with respect to the MgO lattice to separate the energy levels via Zeeman splitting.

We study TiH species without any observable nuclear spin adsorbed on bridge sites with different orientations with respect to the external magnetic field, as sketched in Fig. 1a. TiH on MgO has been shown to be an effective spin-1/2 particle with an anisotropic g-factor [22,23]. In agreement with these studies, we observe different ESR resonance frequencies for the two species: for the spin $S_v$ vertically oriented TiH species (blue) we find a g-factor $g_v$ = 1.75 while for the spin $S_h$ of the horizontal TiH (green) we find $g_h$ = 1.95 (supplementary sections S1 and S2).

Figures 1b and c demonstrate how we use the effective magnetic field emanating from the tip on one of the two atoms to tune the level of entanglement between the spins [10, 17]. If the two spins experience the same effective Zeeman splitting, they precess at identical Larmor frequencies resulting in entangled states. Since we want to reach entanglement at a finite tip field, the two spins need to be inherently detuned in absence of the tip. For this reason, we build heterodimers out of vertically and horizontally oriented TiH species (Fig. 1b, see supplementary section S3 for details).

The dimers are engineered to have a spacing of 1.3 nm, corresponding to a coupling strength in the order of tens of MHz. This coupling strength is chosen to ensure that the dynamics of the local magnetization are slow enough to be probed by our experimental setup, which is limited to ~5 ns pulses, but still faster than the ~100 ns decoherence time of TiH dimers [17]. At this distance, the atoms interact both via exchange and dipolar interactions. The Hamiltonian of the system can be written as (supplementary section S4):

$$\mathcal{H} = (J+2D)S_v^z S_h^z + (J-D)(S_v^x S_h^x + S_v^y S_h^y) - \mu_B B_{ext}(g_v S_v^z + g_h S_h^z) - \mu_B B_{tip} g_v S_v^z$$

where $J$ and $D$ are the exchange and dipolar coupling parameters, respectively. The two last terms account for the Zeeman splitting due to the external ($B_{ext}$) and effective tip ($B_{tip}$) fields, which for simplicity we assume to be aligned.

We separate the exchange and dipolar contributions by performing the experiment on two heterodimers as sketched in Figs. 2a (dimer A) and b (dimer B). The two dimers are equidistant, yielding identical exchange couplings. However, as they are oriented at different angles with respect to the external field, their dipolar coupling strengths differ. This is confirmed by ESR measurements performed on top of the vertically oriented TiH of each dimer with the tip well away from the tuning point (Fig. 2a and b). In this situation, the $S_x S_x$ and $S_y S_y$ components of the coupling (Eq. 1) average out over time as the spins precess with different Larmor frequencies. The resulting coupling, being mediated through the $S_z S_z$ terms only, is effectively Ising-like. Due to the composition of the eigenstates and because ESR can only flip the spin underneath the tip [24], only transitions I and II are observed (see Fig. 1c). The measured splitting between these two ESR resonances corresponds to $J + 2D$, and thus, is different for the two heterodimers (See Fig. 2a and b).

In order to probe the full energy level diagram of Fig. 1c and identify the exact tuning point for maximal entanglement, we perform ESR measurements at various tip heights for each dimer (Figs. 2c and 2d). We observe two sets of peaks that, upon tip approach, shift together and broaden due to decoherence effects [25]. These two sets of resonances can be assigned to transitions I & II and III & IV in Fig. 1c. Away from the tuning point, only one of these pairs is observed: transition I & II before the tuning point and transition III & IV after it. As the energy eigenstates become more entangled near the tuning point, all four transitions become accessible. Due to the opposite signs of the dipolar coupling contributions, the two dimers show slightly different behaviors: transitions II and III intersect twice for dimer A while they stay apart for dimer B. We find $J = 67 \pm 2$ MHz, $D = 2 \pm 1$ MHz for dimer A and $D = -15 \pm 1$ MHz for dimer B (supplementary section S5).

We now arrive at the second stage of the experiment where we measure the free time evolution of the spins. We use a pump probe scheme to excite and measure the spin state of the atom underneath the tip for various degrees of entanglement. When the tip height is far away from the tuning point, the pump probe experiments show the onset of an exponential decay similar to the decay signal of a single excited spin (Figs. 2e and f, top and bottom curves) [9]. By contrast, when tuned, we observe a clear oscillation with a frequency of $64 \pm 1$ MHz for dimer A and $84 \pm 1$ MHz for dimer B. We attribute these oscillations to the flip-flop interaction of strength $J - D$ between the two atoms in the dimer.

The dynamics of the flip-flop interaction can be well understood by describing the time evolution of the combined density matrix of the two spins within a dissipative Bloch-Redfield framework [26,27], which accounts for the uncorrelated electron baths in sample and tip (Fig. 3a, see supplementary section S6 for details). In Figs. 3b-e we show density matrices in the energy basis obtained by numerical simulation for a perfectly tuned dimer at different moments in time after the pump pulse.

During the pump pulse, the system is pushed into a coherent superposition of its excited states (Fig. 3b). These add up to a net $|\downarrow\uparrow\rangle$ magnetization (where the left arrow corresponds to the spin underneath the tip), as a result of spin pumping [21]. This net magnetization is reflected in the off-diagonal terms, which correspond to the coherence between the $|-\rangle$ and $|+\rangle$ states. Immediately after the pulse, the off-diagonals begin to oscillate between positive and negative values (Figs. 3c,d), giving rise to the observed periodicity in the magnetization (see inset of panel a). Due to interaction with the electron baths, the oscillations decay over an effective decoherence time and eventually the populations evolve back towards thermal equilibrium (Fig. 3e). We estimate the decoherence time to be 60 ns and 130 ns for the relaxation time (supplementary section S7).

We now proceed to the effect of detuning on the flip-flop oscillations. For this purpose, we perform pump-probe experiments at different current setpoints around the tuning point for each dimer (Fig. 4a and b). As expected, we find a marked decrease in oscillation amplitude for increased detuning. Depending on the microscopic tip apex, we observe a small difference in tuning height between ESR (Fig. 2) and pump probe measurements (Fig. 4). We attribute this discrepancy to the temperature dependence of the spin polarized current as the two different experiments were carried out at different temperatures (supplementary section S8).

To gain insight into the effect of detuning on the spin dynamics, we map the effective two-level system of the inner 2×2 matrix of the density matrix onto a Bloch sphere. For clarity, the axes of the sphere are fixed to be the energy eigenstates of the fully tuned case, while the projected spin state evolution is plotted for different levels of detuning (Fig 4c-f). As can be seen from the density matrices in Fig. 3, the spin state always has components outside the inner 2x2 matrix, meaning that the projection in Fig. 4 never reaches the surface of the sphere.

When the dimer is in tune (Fig. 4c), the state moves fully within the vertical plane of the Bloch sphere, making maximal flip-flops between $|\uparrow\downarrow\rangle$ and $|\downarrow\uparrow\rangle$. With increasing detuning, the axis around which the state rotates moves as the eigenstates of the system gradually tilts towards the vertical. The difference between the projected maxima and minima of the oscillation onto the vertical axis gets smaller and thus the oscillation amplitude decreases, consistent with our experimental observations.

The observed flip-flop frequency remains constant as a function of detuning. This is surprising, as the energy splitting is supposed to increase away from the tuning point, causing an increase in the frequency (see supplementary section S9). Since the in-plane anisotropy of the g-factor indicates a partially unquenched orbital moment resulting from the crystal-field symmetry of the bridge sites, as previously observed for the out-of-plane direction on TiH species on oxygen binding sites [22], the observed discrepancy may be related to this orbital moment.

In conclusion, by combining the energy resolution of ESR-STM and the time resolution of pump-probe spectroscopy, we have demonstrated a new experimental procedure enabling the observation of the free coherent evolution of coupled atomic spins. As the dynamic processes are initialized by means of a coherence-preserving pulse in the tunneling current, our results provide novel insight into the locality of electron-spin scattering: only the spin directly underneath the tip is affected, irrespective of its global quantum state. In conjunction with the recent demonstration of pulsed ESR-STM, our technique offers pathways towards coherent manipulation of extended atomic spin arrays. The ability to perform a very local, nearly instantaneous, coherent spin flip inside an extended spin lattice constitutes an essential building block for advances in spintronic engineering as well as studies into the propagation of spin waves.

**Acknowledgments:** We thank V. Kornich for scientific discussions.

**Funding:** The authors acknowledge support from the Netherlands Organisation for Scientific Research (NOW Vici Grant VI.C.182.016) and from the European Research Council (ERC Starting Grant 676895 "SPINCAD"). M.T. acknowledges support by the Heisenberg Program (grant no. TE 833/2-1) of the German Research Foundation.

**Author contributions:** L.M.V., L.F. and R.R. performed the experiments. L.M.V., L.F., R.R., D.C. and J.G. implemented and optimized the experimental techniques. L.M.V and L.F. analyzed the experimental data. R.B. and M.T. developed the dissipative Bloch-Redfield model and performed the simulations. L.M.V., R.B. and A.F.O. designed the experiment. L.M.V., L.F., R.B., M.T. and A.F.O. wrote the manuscript, with input from all authors.

**Competing interests:** The authors declare no competing interests.

**Data and materials availability:** All data presented in this paper is publicly available via digital object identifier (DOI) 10.5281/zenodo.4467820.

# Figures

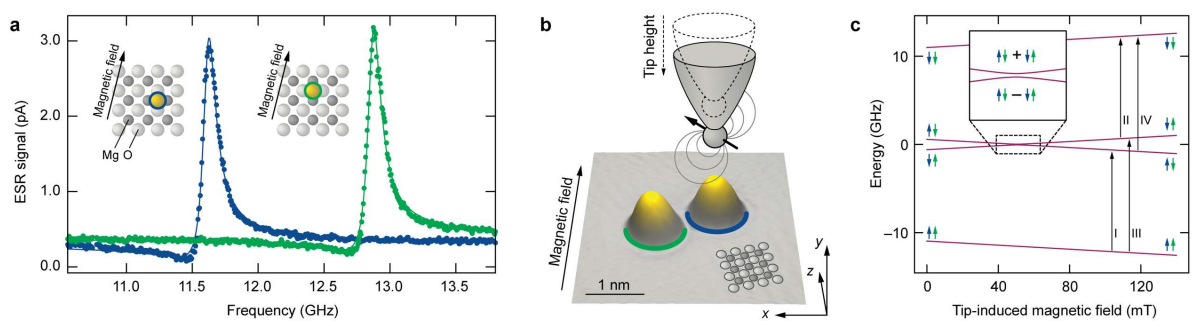

**Fig. 1. Tuning the eigenstates of a TiH dimer using tip-induced magnetic field.** (a) ESR measurements of single TiH adsorbed onto a vertical (blue) and a horizontal (green) bridge site ($T$ = 1.5 K, $V_{RF}$ = 30 mV, $I$ = 50 pA, $V_{DC}$ = 60 mV, $B_{ext}$ = 450 mT). (b) STM topography of a TiH dimer with MgO lattice indication and schematic demonstrating tuning of the tip field. (c) Calculated energies and eigenstate compositions as function of tip field. An avoided crossing occurs at the point where the tip field compensates the g-factor difference. Roman numerals indicate the four possible ESR transitions.

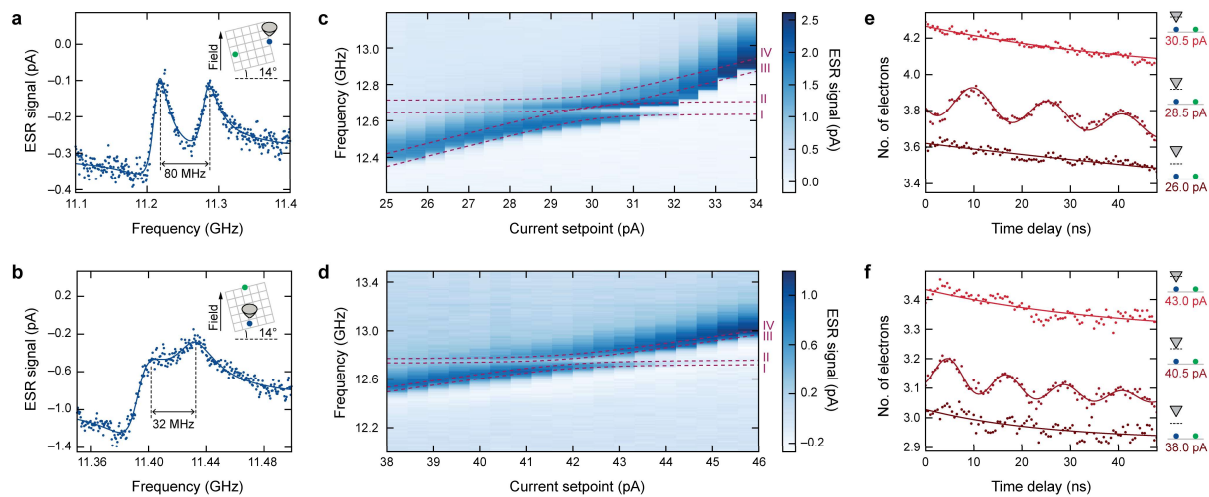

**Fig. 2. Measurement of free coherent evolution at the tuning point.** (a) and (b) ESR measurements on dimer A and dimer B ($T$ = 1.5 K, $V_{RF}$ = 50 mV, $I$ = 10 pA, $V_{DC}$ = 60 mV). Insets: schematic drawings of the dimer placement on the MgO lattice. (c) and (d) ESR measurements at various tip heights, showing an avoided crossing at the tuning point ($T$ = 1.5 K, $V_{RF}$ = 50 mV, $V_{DC}$ = 60 mV). Dashed lines: guides to the eye marking ESR transitions. (e) Pump-probe measurements on dimer A, above, below and at the tuning point ($T$ = 400 mK, setpoint voltage 60 mV, pulse width 7 ns). (f), same for dimer B (pulse width 5 ns). All pump-probe experiments use +100 mV pump and −100 mV probe pulses.

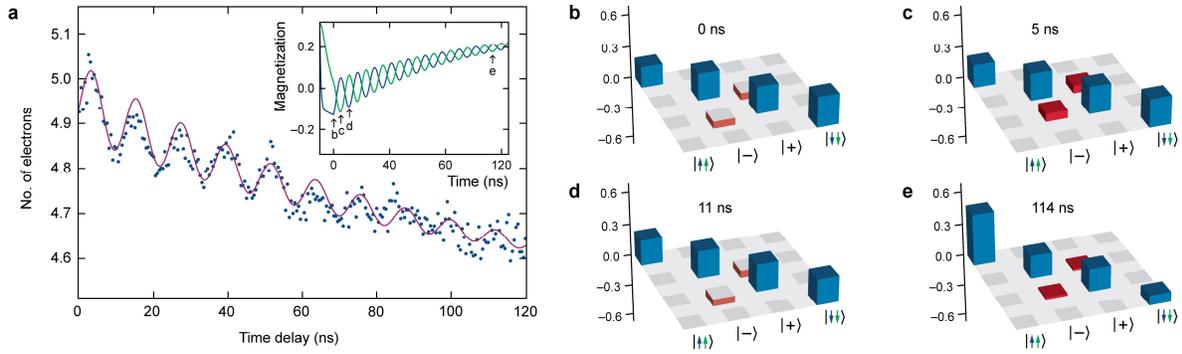

**Fig. 3. Decoherence of the flip-flop oscillation.** (a) Pump-probe measurement on dimer B showing the decay in amplitude of the oscillations ($T$ = 400 mK, setpoint 40.5 pA, 60 mV, pulse width 7 ns). Solid line: calculated pump-probe signal. Inset: calculated magnetizations $\langle S_v^z \rangle$ (blue) and $\langle S_h^z \rangle$ (green); the origin of the time axis is set to coincide with the end of the pump pulse. (b-e) Density matrices at different times after the pump pulse (see inset of panel a). Off-diagonal elements are marked red for clarity.

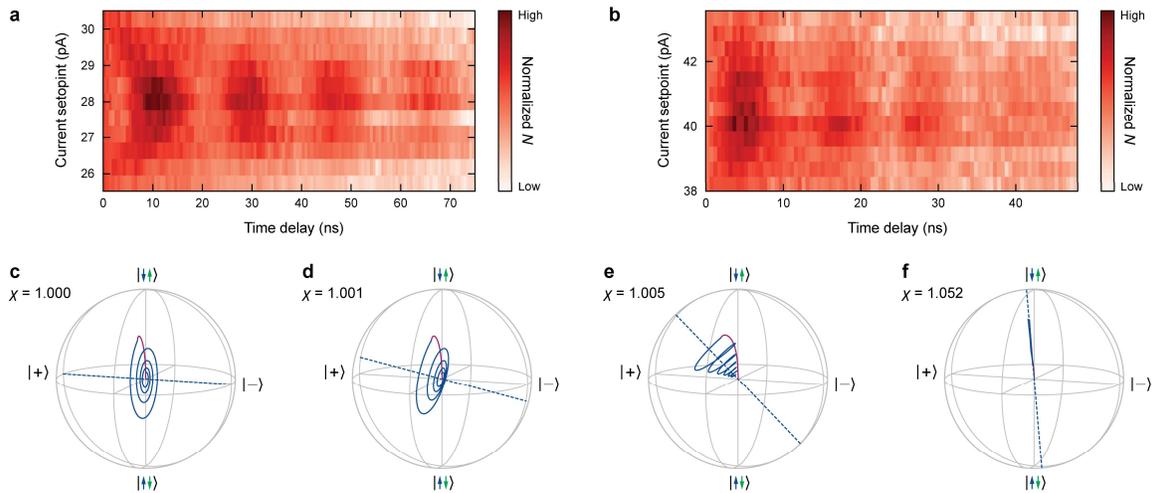

**Fig. 4. Flip-flop oscillations as function of detuning.** (a) and (b) Pump-probe measurements on dimer A and B at various tip heights (parameters same as in Fig. 2). (c-f) Bloch sphere representations of the state evolution at increasing levels of detuning, as indicated by the ratio $\chi = g_v(B_{ext} + B_{tip}) / g_h B_{ext}$. The Bloch spheres show the reduced state space between the $|\uparrow\downarrow\rangle$ and $|\downarrow\uparrow\rangle$ states. The excitation process due to the pump pulse is plotted in red, while the subsequent free evolution of the spin state is plotted in blue.

# Supplementary Materials for

# Free coherent evolution of a coupled atomic spin system initialized by electron scattering

Lukas M. Veldman, Laëtitia Farinacci, Rasa Rejali, Rik Broekhoven, Jérémie Gobeil, David Coffey, Markus Ternes, Alexander F. Otte*.

Correspondence to: a.f.otte@tudelft.nl

**Materials and Methods**

The experiments were performed in a commercial Unisoku USM-1300 ultra-low temperature STM. The sample preparation was done in situ: after cleaning the Ag(100) crystal by Ar sputtering and subsequent annealing, bilayer MgO islands were grown by evaporating Mg at $10^{-6}$ mbar $O_2$ pressure and 650 K sample temperature. We deposited single Fe and Ti atoms onto the sample by means of e-beam evaporation. We used a W tip that we sharpened by indenting it into the Ag until we obtained scanning image quality. All STM topographies were made in constant current mode. All DC bias values are reported with respect to the sample.

For ESR measurements, we used a Keysight MXG N5183B to deliver the RF signal to the junction. The signal was combined with the DC bias voltage via a bias-tee (Tektronix PSPL5542) and applied to the tip (See Fig. S1a). The RF signal was chopped at a few hundred Hz for lock-in detection [11]. Measurements were done in constant-current mode after tracking the center of the atom under investigation. Because of heating due to RF losses in the cabling, we performed all ESR measurements at 1.5 K. In Fig. S1b we show an ESR measurement of a single TiH atom on bilayer MgO. All ESR data are fitted using Fano functions as defined in [25] to determine the resonance frequency. Since we are only interested in the location of the ESR resonance, we do not use the more precise fitting function of Ref. [28] because of the increased number of parameters.

For pump probe experiments we used an Agilent 81110A pulse generator to produce pulse trains. By means of a Mini Circuits ZYSWA-2-50DR switch we connected either the pulse generator or the DC bias to the tip (Fig S1c). For lock-in detection, we cycled through two different pulse train at several hundreds of MHz: one consisting of only pump pulses and one of pump and probe pulses [9]. Pulse height calibration was done by choosing a setpoint using the DC bias, switching to the pulse generator outputting a pulse train with a certain duty cycle and comparing the current to the initial setpoint. We adjust the pulse height to match the current expected from the duty cycle. For optimal signal-to-noise ratio, we rewired every time we switched between the setups of Fig. S1a and Fig. S1c for the different experiments. In Fig. S1d we show pump probe measurements on a single $Ti_B$ which exhibits a lifetime of 141 ns.

## Supplementary Text

Section S1: Sample characterization

We show in Fig. S2a a topography of the sample after co-evaporation of Fe and Ti atoms. The two atomic species have different apparent heights and can be unambiguously identified by recording d$I$/d$V$ spectra above their center. As in the main text, the spectra are recorded with an external magnetic field of 450 mT and a spin polarized tip. Fe atoms on MgO have a total spin of $S$ = 2 and show two pairs of steps at ±1 mV and ±14 mV (orange curve in Fig. S2b) manifesting inelastic spin excitations [11, 29]. A reference spectrum taken above the MgO island is shown in grey. TiH atoms on MgO are effective spin $S$ = 1/2 particles and their apparent height and d$I$/d$V$ spectra depends on their adsorption site [16]. When adsorbed on an O-site, the TiH atoms have a relatively low apparent height and show an inelastic spin excitation around the Fermi energy (red curve in Fig. S2c) as well as an abrupt increase of the conductance around ±90 mV (red curve in Fig. S2d) whose origin may be an orbital excitation [22]. TiH atoms adsorbed on a bridge site (i.e. in-between two O atoms) show a larger apparent height and a similar spin excitation around the Fermi energy (blue curve in Fig. S2c) but no sign of orbital excitation on a broader energy range (blue curve in Fig. S2d).

Section S2: Determination of the g-factors

We can determine the different g-factors of the horizontal and vertical bridge adsorption sites of TiH atoms by recording ESR spectra as a function of set-point current [17]. For both the horizontal bridge site TiH$_h$ (Fig. S3a) and the vertical bridge site TiH$_v$ (Fig. S3b), the resonance shifts to higher frequencies upon tip approach indicating that the effective tip magnetic field is (at least partially) aligned with the external magnetic field. We extract the resonance positions by fitting each spectrum with a Fano function (red curves in Fig. S3a and b) and plot their evolution as a function of setpoint current in Fig. S3c. We observe a linear shift of the resonance positions indicating that we are in a regime where the exchange interaction between tip and TiH atom dominates over the dipolar one [30]. We can thus express the position of the resonance frequency in ESR as a function of setpoint current $I$ as was done in [17]:

$$f = \frac{g\mu_B}{2\pi\hbar}(B_{\text{tip}} + B_{\text{ext}}) = E + aI, \tag{S1}$$

where $E$ and $a$ are parameters that we obtain by fitting a line to the data points in Fig. S3c.

The parameters of the fits are shown in Table 1 where the error bars correspond to the standard deviations of the fits (red lines). The g-factors are deduced from $E$ and we obtain for the vertical TiH: $g_v$ = 1.740 ± 0.006 (corresponding to an effective magnetic moment $\mu_v$ = 0.87± 0.003 $\mu_B$) and for the horizontal species: $g_h$ = 1.953 ± 0.003 ($\mu_h$ = 0.98 ± 0.002 $\mu_B$). Knowing the g-factors we can relate the changes in current setpoint to effective tip field variations: a variation of the current by 1 pA leads to a change in magnetic field of ∼ 2.2 mT for TiH$_h$ and ∼ 2.1 mT for TiH$_v$, indicating a small tilt of the tip field with respect to the MgO lattice. Finally, in Fig. S3d, we can relate the current variations to changes in tip height after opening the feedback with $V_{\text{bias}}$ = 60 mV and $I$ = 10 pA (a negative offset corresponds to a tip approach toward the atom). The forward and backward traces fall on top of each other ensuring the stability and repeatability of the experiment, with contact reached at ∼ −390 pm. A fit of the exponential dependence of the current with distance for the low conductance regime (blue dashed curve in Fig. S3d) gives $I \propto \exp(-d/d_0)$, where $d_0$ = 54.05 ± 0.05 pm.

Section S3: Dimer configuration

We construct heterodimers of TiH atoms adsorbed on bridge sites at different angles with the external magnetic field. The exact position of the atoms with respect to the underlying MgO lattice is determined by obtaining atomic resolution of a MgO island (see Fig. S4a). In this image only one of the two sub-lattice (Mg or O) of the MgO layer is visible: the lattice parameter is $\sim 2.87$ Å (i.e. $A_{Ag}/\sqrt{2}$, where $A_{Ag}$ is the lattice constant of Ag(100)) [16]. The adsorption sites of the atoms refer to their position with respect to the O-lines of the MgO lattice, which we draw in red in Fig. S4. Once the relative orientation of the MgO lattice is determined, we identify the atomic positions by anchoring the lattice at Fe atoms (marked in orange in Fig. S4b and c), which adsorb on O-sites [11, 29]. As one can see in Fig. S4b and c, the two dimers presented in the main text have the exact same spacing but with a 102° angle between them. The horizontal and vertical nature of the bridge adsorption sites for the TiH atoms can be furthermore confirmed by ESR measurements (see main text).

Section S4: System Hamiltonian

This section describes the Hamiltonian of a dimer consisting of hydrogenated Ti atoms taking into account the effective magnetic field $\mathbf{B}_{\text{tip}}$ emanating from the tip. This effective field arises from exchange interaction between the Ti below the tip and the tip itself. Because in our situation this interaction is weak compared to the Zeeman energy of the external field $\mathbf{B}_{\text{ext}}$, it does not significantly alter the ground-state orientation of the spins, which we assume to be aligned with $\mathbf{B}_{\text{ext}}$. Since also the coupling between the Ti atoms themselves is weak compared to the Zeeman energy resulting from $\mathbf{B}_{\text{ext}}$, the Hamiltonian can be further simplified using a secular approximation [31]. Under these conditions the general form of the Hamiltonian is [16, 17]:

$$\widehat{\mathcal{H}} = -\mu_B \left( B_{\text{ext}} + B_{\text{tip}} \right) g_v \hat{S}_v^z - \mu_B B_{\text{ext}} g_h \hat{S}_h^z + J \hat{\mathbf{S}}_v \cdot \hat{\mathbf{S}}_h + D \left( 3 \hat{S}_v^z \hat{S}_h^z - \hat{\mathbf{S}}_v \cdot \hat{\mathbf{S}}_h \right) \quad (S2)$$

where $z$ is chosen to align with $\mathbf{B}_{\text{ext}}$ and $\mu_B$ is the Bohr magneton. The first two terms are the Zeeman splitting of the respective atoms. The third term is the exchange coupling with strength $J$ between the atoms and the last term is the dipolar coupling, the strength $D$ of which is dependent on the angle between $\mathbf{B}_{\text{ext}}$ and the vector separating the two atoms. $\hat{\mathbf{S}}_v$ is the spin operator of the atom below the tip and $g_v$ is the $z$-component of the corresponding anisotropic g-factor; $\hat{\mathbf{S}}_h$ and $g_h$ belong to the other atom. The detuning, i.e. the difference between the Zeeman splittings of the two spins, is given by $\mu_B \left( g_v (B_{\text{ext}} + B_{\text{tip}}) - g_h B_{\text{ext}} \right)$. When the Hamiltonian is rewritten as in (S3) it becomes apparent that the exchange and dipolar coupling between the atoms causes an ESR splitting of $J + 2D$ and a transverse flip-flop coupling of $J - D$.

$$\widehat{\mathcal{H}} = (J + 2D) \hat{S}_v^z \hat{S}_h^z + (J - D)\left( \hat{S}_v^x \hat{S}_h^x + \hat{S}_v^y \hat{S}_h^y \right) - \mu_B B_{\text{ext}} \left( g_v \hat{S}_v^z + g_h \hat{S}_h^z \right) - \mu_B B_{\text{tip}} g_v \hat{S}_v^z \quad (S3)$$

We choose the combined eigenstates of $\hat{S}_v^z \hat{S}_h^z$ ($|\uparrow\uparrow\rangle, |\downarrow\uparrow\rangle, |\uparrow\downarrow\rangle, |\downarrow\downarrow\rangle$), known as the Zeeman product states, as computational basis. Within that basis the eigenstates and corresponding eigenenergies of the dimer Hamiltonian are:

| $n$ | Eigenstate $|n\rangle$ | Eigenenergy $E_n$ |
|---|---|---|
| 3 | $|\downarrow\downarrow\rangle$ | $\frac{1}{4}(J+2D) + \frac{1}{2}\mu_B \left( g_v \left( B_{ext} + B_{tip} \right) + g_h B_{ext} \right)$ |
| 2 | $\eta \geq 0 : \cos\left(\frac{\theta}{2}\right)|\uparrow\downarrow\rangle + \sin\left(\frac{\theta}{2}\right)|\downarrow\uparrow\rangle$<br>$\eta \leq 0 : \sin\left(\frac{\theta}{2}\right)|\uparrow\downarrow\rangle + \cos\left(\frac{\theta}{2}\right)|\downarrow\uparrow\rangle$ | $-\frac{1}{4}(J+2D) + \frac{1}{2}\sqrt{(J-D)^2 + \left(\mu_B \left( g_v \left( B_{ext} + B_{tip} \right) - g_h B_{ext} \right)\right)^2}$ |
| 1 | $\eta \geq 0 : \sin\left(\frac{\theta}{2}\right)|\uparrow\downarrow\rangle - \cos\left(\frac{\theta}{2}\right)|\downarrow\uparrow\rangle$<br>$\eta \leq 0 : \cos\left(\frac{\theta}{2}\right)|\uparrow\downarrow\rangle - \sin\left(\frac{\theta}{2}\right)|\downarrow\uparrow\rangle$ | $-\frac{1}{4}(J+2D) - \frac{1}{2}\sqrt{(J-D)^2 + \left(\mu_B \left( g_v \left( B_{ext} + B_{tip} \right) - g_h B_{ext} \right)\right)^2}$ |
| 0 | $|\uparrow\uparrow\rangle$ | $\frac{1}{4}(J+2D) - \frac{1}{2}\mu_B \left( g_v \left( B_{ext} + B_{tip} \right) + g_h B_{ext} \right)$ |

The eigenstates are a function of $\theta$ which in turn is a function of $\eta$, the ratio between the detuning and the flip-flop coupling.

$$\theta = \arctan\left(\frac{1}{|\eta|}\right) \quad (S4)$$

$$\eta = \frac{\mu_B \left( g_v \left( B_{ext} + B_{tip} \right) - g_h B_{ext} \right)}{J - D} \quad (S5)$$

Two important limits can be distinguished. When the spins are perfectly tuned, so $\eta = 0$, the eigenstates are singlet-triplet states separated by the flip-flop coupling $J - D$. In contrast, when the atoms are far detuned, so $|\eta| \gg 1$, the eigenstates approach the Zeeman product states separated by the detuning.

Section S5: Determination of coupling parameters

We detail in this section how we estimate the values of the exchange coupling parameter $J$, and dipolar coupling parameter $D$ for the two dimers presented in the main text. As mentioned in section S3, the two dimers have the exact same spacing, leading to identical exchange coupling strength but different orientations with respect to the external field so that their dipolar couplings differ. In particular, as one can see from Fig. S5a, we expect $D_A > 0$ since the dipolar coupling favors antiferromagnetism for dimer A and $D_B < 0$ corresponding to a favored ferromagnetic dipolar coupling for dimer B (Fig. S5b). The dependence of the energy levels on the $J$ and $D$ parameters is given by the eigenenergies as given in Section S4. In ESR we measure transitions between these energy levels and we define:

$$\begin{aligned} f_I &= \frac{1}{2\pi\hbar}(E_1 - E_0) \\ f_{II} &= \frac{1}{2\pi\hbar}(E_3 - E_2) \\ f_{III} &= \frac{1}{2\pi\hbar}(E_2 - E_0) \\ f_{IV} &= \frac{1}{2\pi\hbar}(E_3 - E_1). \end{aligned} \tag{S6}$$

In particular we obtain that the frequency difference between the first and second resonances and the third and fourth resonances does not depend on the tip field:

$$\begin{aligned} \Delta f &= f_{IV} - f_{III} = f_{II} - f_I \\ &= \frac{1}{2\pi\hbar}(E_3 - E_2 - E_2 + E_0) \\ &= J + 2D. \end{aligned} \tag{S7}$$

We determine this frequency splitting by recording a high resolution ESR spectrum with the tip far away to reduce decoherence effects and subsequent broadening of the peaks. By fitting the spectra by the sum of two Fano functions (see Fig. 2a of the main text) we obtain:

$$\begin{aligned} J + 2D_A &= 68.71 \pm 0.25 \text{MHz} \\ J + 2D_B &= 37.48 \pm 0.64 \text{MHz}, \end{aligned} \tag{S8}$$

where the error bars correspond the standard deviations of the fits.

The flip-flop transition observed in the pump-probe measurements directly correlates to the energy splitting between the first and second eigenstates:

$$f_{\text{flip-flop}} = \frac{1}{2\pi\hbar}\sqrt{[(g_v - g_h)\mu_B B_{\text{ext}} + g_v \mu_B B_{\text{tip}}]^2 + [J - D]^2}. \tag{S9}$$

The oscillations are observed when the two spins are tuned so that the first term in the square root equals zero. Fitting the curves of Fig. 2e and f of the main text by a damped sinusoid (see section S8) gives:

$$|J - D_A| = 64.26 \pm 0.35 \text{MHz}$$
$$|J - D_B| = 83.1 \pm 0.81 \text{MHz,}$$
(S10)

where the error bars are the standard deviations of the fits.
From equations (S8) and (S10), we obtain the following estimates:

$$J = 67 \pm 2 \text{MHz}$$
$$D_A = 2 \pm 1 \text{MHz}$$
$$D_B = -15 \pm 1 \text{MHz.}$$
(S11)

In the model used in the main text, we assume the ground-state orientations of the spins to be aligned with **B**$_{ext}$. The obtained values for the dipolar coupling parameters allow us to estimate to what extent this assumption is correct. The dipolar coupling parameter can be expressed as a function of the angle $\Theta$ between the ground-state orientations of the spins and the vector *r* between the two atoms: $D = \frac{D_0}{2}(1 - 3\cos^2 \Theta)$, where $D_0$ is a constant that does not depend on the geometry of the dimer. Since we know the relative orientation of the dimers with respect to each other we can calculate the ratio $D_B/D_A$ as a function of the angle $\alpha$ that the spins in the ground state make with the lattice vector of MgO (see Fig. S5a). The results are plotted in Fig. S5c where the black dashed lines corresponds to the value obtained from the experiment (the grey area representing the error bars) and the red dashed lines show the result obtained when the spins in the ground state are fully aligned with the external magnetic field ($\alpha = 14°$). The small deviation indicates a non-perfect ground-state alignment of the spins with respect to external field. This is expected for an anisotropic g-factor (in this case, the deviation differs for the two species) and can also be influenced by the presence of the spin-polarized tip.

We can ensure that the values obtained for *J*, $D_A$ and $D_B$ correctly reproduce the behavior of the ESR measurements around the tuning point as shown in Fig. 2c and d of the main text. We calculate the dependence of the ESR resonances as a function of tip field using equations (S1) and then use the analysis performed in section S2 to convert the tip field to an effective tunneling current. We find a good agreement with the measurements if we include a current and frequency offset to the calculations: the guide to the eye of Fig.2c (Fig.2d) is obtained using the *J*, $D_A$ and $D_B$ values of eq. (S11), a current offset of 7.6 pA (19.5 pA) and a frequency offset of 0.38 GHz (0.45 GHz). See Section 8 for a discussion on the possible origins for these offsets.

The qualitative behavior of the ESR peaks at the tuning point can also be understood by considering the frequency difference between the second and third ESR resonances. Using equations (S6) we have:

$$f_{III} - f_{II} = \frac{1}{2\pi\hbar}(2E_2 - E_0 - E_3)$$
$$= -J - 2D + \sqrt{[(\gamma_1 - \gamma_2)\hbar B_{ext} + \gamma_1 \hbar B_{tip}]^2 + [J - D]^2},$$
(S12)

since $J > D$, we have at the tuning point:

$$f_{III} - f_{II} = -3D \tag{S13}$$

As a result, when $D > 0$, we have $f_{III} > f_{II}$ at the tuning point and the resonances cross twice upon tip approach, as is the case for dimer A. By contrast, when $D < 0$, the two resonances stay apart from each other as observed for dimer A.

The guides to the eye shown in Fig. 2c and d of the main text were produced by calculating the resonances shown in equations (S6) as function of tip field. In order to match the calculation with the experiment, we found that we needed constant offsets in both energy and setpoint axes depending on the microscopic apex of the tip used for measurements.

A better quantitative agreement may be achieved by making the model of Section S4 more realistic. First, we note that we neglected the *x* and *y* components of the tip field while these must be present for driving the ESR resonant transitions [28]. Second, the tip field is assumed to not affect the TiH$_h$ atom of the dimer. Third, we did not separately treat the spin and orbital degrees of freedom and considered the effective anisotropy of the g-factor along the external field rather than along the crystal lattice. Fourth, we assumed that the ground state orientations of the spins are oriented along the direction of the external field. Finally, we use the secular approximation in which the orientations of the dipolar fields are fixed in space.

Section S6: Bloch-Redfield equation for Kondo interaction with the surface:

**Kondo interaction**
As a result of interaction with its environment the ideal dimer magnetization oscillation (see Section S9) vanishes over time. Leading order is the Kondo interaction with the many itinerant electrons of the reservoirs in substrate and tip [12]. These reservoirs can be described in second quantization as:

$$\widehat{\mathcal{H}}_l = \sum_{\mathbf{k},\sigma} \epsilon_{lk\sigma} \hat{a}^\dagger_{l\mathbf{k}\sigma} \hat{a}_{l\mathbf{k}\sigma} \tag{S14}$$

whereby $l$ indicates the substrate ($l = s$) and tip electrons ($l = t$) at an energy $\epsilon$, momentum $k$, and spin $\sigma$, respectively.

We model the Kondo-like interactions as Heisenberg point contact exchange interactions with isotropic coupling strength $J_{l,i}$ between the individual adatom spins $\hat{\mathbf{S}}_i$ and the electrons of the reservoir $l$:

$$\widehat{\mathcal{H}}_{\text{Kondo},l,i} = J_{l,i} \hat{\mathbf{S}}_i \cdot \hat{\mathbf{s}}_l \tag{S15}$$

We approximate the Bloch waves of the itinerant surface electrons in the dimer spin system by plane waves. Note that scattering at the two, spatially separated impurity sites can be correlated leading to interference effects due to phase differences [32]. To account for such effects, the operators $\hat{\mathbf{s}}_s = (\hat{s}^x_s, \hat{s}^y_s, \hat{s}^z_s)^T$ in equation (S15) have to be described as:

$$\hat{s}_s^\alpha = \sum_{\mathbf{k},\mathbf{k}',\sigma,\sigma'} e^{i(\mathbf{k}'-\mathbf{k})\cdot \mathbf{r_i}} \hat{a}^\dagger_{s,\mathbf{k},\sigma} \frac{\tau^\alpha_{\sigma,\sigma'}}{2} \hat{a}_{s,\mathbf{k}',\sigma'} \tag{S16}$$

where $\tau^\alpha$ are the Pauli spin matrices and $\mathbf{r_i}$ is the position vector of the $i$-th spin. However, due to the relatively large separation of the spin sites $|\mathbf{r_1} - \mathbf{r_2}|$ which exceeds significantly the Fermi-wavelength and the large angular degree of freedom of the participating electrons these correlation effects are small in our case and will be neglected in the following. This allows to treat the scattering on the two sites independently and to neglect the phase factor so that the spin operators on tip and sample can be described by:

$$\hat{s}_l^\alpha = \sum_{\mathbf{k},\mathbf{k}',\sigma,\sigma'} \hat{a}^\dagger_{l,\mathbf{k},\sigma} \frac{\tau^\alpha_{\sigma,\sigma'}}{2} \hat{a}_{l,\mathbf{k}',\sigma'} \tag{S17}$$

We furthermore simplify the problem by assuming in the small energy range of interest around the Fermi energy, for tip and sample continuous and energetically flat density of states $\varrho_l(\epsilon) = \sum_{\sigma,k(\epsilon),k'(\epsilon)} \delta_{k,k'} \langle \hat{a}^\dagger_{l,\mathbf{k},\sigma} \hat{a}_{l,\mathbf{k}',\sigma} \rangle \equiv \varrho_l$. To account for any spin polarization in the tip's electron reservoir we use a spin density matrix $\rho_t = \frac{1}{2}(\hat{1} + \mathbf{n}\cdot\hat{\boldsymbol{\tau}})$, with $\mathbf{n}$ as the direction and $0 \leq |\mathbf{n}| \leq 1$ the amplitude of the polarization in the coordinate system. With this convention, the spin polarization becomes identical to the relative imbalance between majority ($\varrho_\uparrow$) and minority ($\varrho_\downarrow$) spin densities $\eta = |\mathbf{n}| = (\varrho_\uparrow - \varrho_\downarrow)/(\varrho_\uparrow + \varrho_\downarrow)$.

**Bloch-Redfield equation**

We continue to solve the time evolution of the dimer system including just the Kondo interactions with surface and tip. We use second order perturbation theory and apply Markovian statistics, since surface and tip are large reservoirs with short correlation times. Then, the Bloch-Redfield master equation can be used to describe the time evolution of the density matrix $\rho$ of the dimer perturbed by its environment. In the energy basis of the dimer the Bloch-Redfield master-equation has the form:

$$\dot{\rho}_{nm}(t) = -i\omega_{nm}\rho_{nm}(t) + \sum_{kj}\left(R_{nmkj} + R'_{nmkj}(t)\right)\rho_{kj}(t) \tag{S18}$$

where $\omega_{nm} = (E_n - E_m)/\hbar$ and $\rho_{nm} = \langle n|\rho|m\rangle$ are the entries of the density matrix.

The first term of the right side of the equation describes the unperturbed state evolution (see Section S9)The second term sums up all the environmentally driven changes of the density matrix by the Bloch-Redfield tensors $R_{nmkj}$ and $R'_{nmkj}$. The interactions with the individual electron baths are described by $R_{nmkj}$ which has the form[26, 27]:

$$R_{nmkj} = \frac{1}{\hbar^2} \sum_{i,l} (J_{i,l})^2 \sum_{\alpha,\beta} \begin{pmatrix} -\delta_{mj} \sum_p \langle n|\hat{S}_i^\alpha|p\rangle\langle p|\hat{S}_i^\beta|k\rangle g_l^{\alpha\beta}(\omega_{pk}) \\ -\delta_{nk} \sum_p \langle j|\hat{S}_i^\alpha|p\rangle\langle p|\hat{S}_i^\beta|m\rangle \left(g_l^{\alpha\beta}(\omega_{pj})\right)^* \\ +\langle n|\hat{S}_i^\beta|k\rangle\langle j|\hat{S}_i^\alpha|m\rangle \left(g_l^{\alpha\beta}(\omega_{nk}) + \left(g_l^{\alpha\beta}(\omega_{mj})\right)^*\right) \end{pmatrix} \quad \text{(S19)}$$

where the first sum counts over the two adatoms and over sample and tip, and $\alpha, \beta$ are the spin degrees of freedom. We assume for the two spins the same coupling strength to the substrate, i.e. $J_{1,s} = J_{2,s} \equiv J_s$, and that only the first spin couples to the tip electrode, i.e. $J_{1,t} = J_t$ and $J_{2,t} = 0$.
The correlation function $g_l^{\alpha\beta}(\omega_{nm})$ is defined as

$$g_l^{\alpha\beta}(\omega_{nm}) = \int_0^\infty dt \langle s_l^\alpha(t) s_l^\beta(0)\rangle e^{-i\omega_{nm}t} \quad \text{(S20)}$$

where the angle brackets indicate the average at thermal equilibrium and the electron spin operators are as defined in (S17). We neglect the imaginary part since it only results in a renormalization of the energy levels [32]. Using Wicks theorem the remaining real part can be evaluated to be equal to [27]:

$$g_l^{\alpha\beta}(\omega_{nm})) = \frac{\hbar}{4} \pi \varrho_l^2 \sum_{\sigma,\sigma'} \frac{\tau_{\sigma,\sigma'}^\alpha \tau_{\sigma',\sigma}^\beta}{4} \iint d\epsilon d\epsilon' f(\epsilon_l)(1 - f(\epsilon_l'))\delta(\epsilon_l - \epsilon_l' - \epsilon_{nm}) \quad \text{(S21)}$$

with $f(\epsilon)$ the temperature dependent Fermi-Dirac distribution and $\delta(\epsilon)$ the Dirac delta function. It has a clear interpretation since it sums over all possibilities for electrons to scatter from an occupied state $\epsilon_l$ to an unoccupied state $\epsilon_l'$ of the reservoir while remaining energy conservation.

To be able to simulate the pump-probe experiments in which a time-dependent bias is applied to the junction, we add additionally the Bloch-Redfield tensor $R'_{nmkj}(t)$ which incorporates bias driven scattering processes between tip and sample $t \rightarrow s$ and sample and tip $s \rightarrow t$ via the spin on site 1, i.e. the Ti$_v$ adatom. In contrast to the scattering processes which begin and end in the same baths, here we include an additional Coulomb-like interaction term $U$ to account for magneto-resistive tunneling effects [21, 33]:

$$\widehat{\mathcal{H}}_{\text{Tunnel}} = \sqrt{\frac{(J_s)^2 (J_t)^2}{(J_s)^2 + (J_t)^2}} \left(\hat{\mathbf{S}}_1 \cdot \hat{\mathbf{s}}_{ts} + \hat{\mathbf{S}}_1 \cdot \hat{\mathbf{s}}_{st} + U\right) \quad \text{(S22)}$$

We calculate the total coupling by assuming that the individual couplings between baths and impurity adding up like serial resistors. The spin operators acting between tip and sample are

$$\hat{s}_{ll'}^\alpha = \sum_{\mathbf{k},\mathbf{k}',\sigma,\sigma'} \hat{a}_{l,\mathbf{k},\sigma}^\dagger \frac{\tau_{\sigma,\sigma'}^\alpha}{2} \hat{a}_{l',\mathbf{k}',\sigma'} \quad \text{(S23)}$$

and the corresponding correlation functions are

$$g_{ll'}^{\alpha\beta}(\omega_{nm})) = \frac{\hbar}{4}\pi\varrho_l\varrho_{l'}\sum_{\sigma,\sigma'}\frac{\tau_{\sigma,\sigma'}^{\alpha}\tau_{\sigma',\sigma}^{\beta}}{4}\iint d\epsilon_l d\epsilon_{l'} f(\epsilon_l)(1-f(\epsilon_{l'}))\delta(\epsilon_l - \epsilon_{l'} - \epsilon_{nm} - V) \tag{S24}$$

with $V$ the applied bias to the tip.

We mimic the lock-in detecting of the pump-probe experiment as show in figure 3 of the main text by simulating an idealized voltage sequence with varying pump-probe delay time (Figure S6A). To reach agreement between simulation and experiment, the voltage during pump and probe pulses has a lower amplitude $|V|=95.3\,\text{mV}$ than the one used in the experiment. Also the length of the pulses is with $t_p = 9.9\,\text{ns}$ slightly longer. Both adjustments compensate for the finite bandwidth of the bias cables and the settling time of the arbitrary waveform generator. Outside the pulse times we assume $V=0$.

For each delay time we then calculate the time dependent current $I(t)$ flowing between tip and sample. Integrating the current over time leads to the average charge carrying electrons per sequence, $n = \int_{-\infty}^{+\infty} dt I(t)/e$ from which we subtract the average electrons crossing the junction during a pump pulse only (see Figure S6B).

We reach best agreement between simulation and experiment when using an effective, bias independent spin-polarization $\eta = 0.314$ in the tip electrode which is aligned with the external field $B = 0.45\,\text{T}$ and a Coulomb scattering parameter $U = \frac{1}{4}$. We furthermore use coupling strengths of $J_s\varrho_s = -0.022$ and $J_t\varrho_t = -0.0048$, which leads to an overall conductivity between tip and sample of $G = 80.2\,\mu\text{S}$ and corresponds well to the setpoint parameters prior the pump-probe measurements of $V = 60\,\text{mV}$ and $I = 40.5\,\text{pA}$.

Figure S6C shows the effect of the pulses on the density matrix $\rho$ for the coupled dimer spin system by tracing out the individual sub-systems and plotting the corresponding $\langle m_z \rangle$ expectation values. At the start of the pump pulse both spins are aligned with the external field. The electrons flowing from sample to tip during the applied pump voltage quickly polarizes the Ti$_v$ system to be anti-parallel aligned to the external field. Also the Ti$_h$ spin which does not rest in the tip-sample junction becomes slowly polarized due to the coupling between both spins. After the pump pulse has ended, the $\langle m_z \rangle$ values of both spins oscillate until the probe pulse current in opposite direction polarizes the Ti$_v$ spin and successively the Ti$_h$ spin back into the direction of the applied field. The current during this probe pulse depend on the $\langle m_z \rangle$ value of the Ti$_v$ spin, as shown in the inset of figure S6}, and is the base for the detection scheme.

Section S7: Lifetime measurements:
In order to obtain an estimate for the effective spin relaxation time $T_1$ for the dimers near the tuning point, we perform pump-probe measurements with 50 ns wide pulses (Fig. S7b). We find an exponential decay when the tip is above the TiH$_h$ (blue) while there is a deviation form exponential behavior for the TiH$_v$ atom (orange) when the tip is at the tuning height. With the tip above the TiH$_h$, the dimer is detuned and the excitation caused by the pump pulse remains on the atom under the tip. When the tip is above the TiH$_v$ however, the dimer is in tune and the excitation is shared between the two atoms. This causes a drop in magnetization measured by the pump pulse up until the point where the spin state is decohered and the tuned and detuned cases become indistinguishable. As a measure for the relaxation time of the spin excitation on the dimer, we fit the exponential decay at the tuning point (feedback opened at 28 pA, 60 mV) on the TiH$_h$; we find $T_1$ = 130 ± 5 ns.

The decoherence time $T_2$ is determined by fitting a damped sinusoid to the data shown in Fig 3a in the main text.

$$Ae^{-t/T_2}\sin(2\pi ft)+Be^{-t/T_1}+C, \tag{S25}$$

where $f$ is the frequency of the oscillations, $T_1$ and $T_2$ characteristic decay times, $C$ an offset, and $A$ and $B$ scaling coefficients for the oscillations and exponential decay, respectively. We find $T_2$ = 59 ± 9 ns and $T_1$ = 101 ± 13 ns, where the error bars are the standard deviations of the fits (Fig. S7a). The smaller T1 could be assigned to the fact that when the dimer is in tune, the excitation is shared between the two atoms and therefore two scattering centers that connect the excitation to the electron bath. We note that the data was recorded only over the onset of the exponential decay (the first 125 ns) and that the definition of $T_1$ as in equation (S21) is simplistic since when the dimer is in tune the system is more complex than an effective two level system. We therefore give the upper limit as determined in Fig. S7b for our estimate in the main text.

Section S8: Temperature dependence
We notice a small difference (∼ 2 pA) between our ESR and pump-probe measurements concerning the set-point current at which the perfect tuning point is reached. We suggest that this is due to the temperature dependence of the TiH$_v$ atom polarization as well as, possibly, a change in the spin polarization of the tip. We consider that the current depends exponentially on the distance between tip and sample $d$ and linearly on the conductance $G$ of the junction ($I = V_{\text{bias}}G\exp(-\alpha d)$). Additionally, this conductance has a spin polarized component:

$$G = G_0(1+\beta\langle S_{\text{TiHv}}\rangle\langle S_{\text{tip}}\rangle), \tag{S26}$$

where $\beta$ scales the spin-polarized versus non spin-polarized contributions to the conductance. Since the exchange field of the tip has the same exponential dependence on $d$ as the current $I$ ( $B_{\text{tip}} = K\exp(-\alpha d)$ ), we can write:

$$B_{\text{tip}} = \frac{K}{G_0(1+\beta\langle S_{\text{TiHv}}\rangle\langle S_{\text{tip}}\rangle)}I. \tag{S27}$$

The coefficient $a$ determined in section S2 is thus dependent on the temperature of the experiment. In Fig. S8 we calculate the magnetization of the atom under the tip when the dimer is in the Boltzmann ground state for different temperatures. We find a significant difference in magnetization between the experimental temperatures: 400 mK and 1.5 K. This may explain the small discrepancy in set point current (2 pA) between the ESR and pump-probe measurements.

Section S9: Ideal flip-flop interaction as function of detuning

We describe now how the oscillations observed in the time evolution of the magnetization of the TiH$_v$ atom are expected to evolve with detuning.

Ideally the dimer state $|\psi\rangle$ is fully initialized in the $|\downarrow\uparrow\rangle$ state by local excitation of the atom underneath the tip. The subsequent unperturbed time evolution of this state is fully governed by the dimer Hamiltonian $\widehat{\mathcal{H}}$.

$$|\psi(t)\rangle = e^{-i\widehat{\mathcal{H}}t}|\downarrow\uparrow\rangle$$

Inserting (S3) we find the dimer state as a function of time:

$$|\psi(t)\rangle = \begin{cases} \left(\sin^2\left(\frac{\theta}{2}\right)a_2 + \cos^2\left(\frac{\theta}{2}\right)a_1\right)|\downarrow\uparrow\rangle + \cos\left(\frac{\theta}{2}\right)\sin\left(\frac{\theta}{2}\right)(a_2-a_1)|\uparrow\downarrow\rangle & \text{for } \eta \geq 0 \\ \left(\cos^2\left(\frac{\theta}{2}\right)a_2 + \sin^2\left(\frac{\theta}{2}\right)a_1\right)|\downarrow\uparrow\rangle + \cos\left(\frac{\theta}{2}\right)\sin\left(\frac{\theta}{2}\right)(a_2-a_1)|\uparrow\downarrow\rangle & \text{for } \eta \leq 0 \end{cases} \quad \text{(S28)}$$

with $E_i$ the eigenenergies of the dimer Hamiltonian and $a_i = \exp(-\frac{i}{\hbar}E_i t)$

We can then find the magnetization of the atom below the tip as:

$$M = \langle\psi(t)|\hat{S}_v^z \otimes \hat{I}_h|\psi(t)\rangle$$
$$= \frac{1}{2}\cos^2(\theta) - \frac{1}{2}\sin^2(\theta)\cos(\omega_{12}t) \quad \text{(S29)}$$

where $\omega_{12} = \dfrac{E_1 - E_2}{\hbar}$, and $\hat{I}_h$ is the identity operator.

We can finally write this result in terms of $\eta$ the ratio between detuning and flip-flop coupling:

$$M = -\frac{1}{2}\left(\frac{\eta^2}{1+\eta^2} + \frac{1}{1+\eta^2}\cos\left(\sqrt{1+\eta^2}\frac{|J-D|}{\hbar}t\right)\right) \quad \text{(S30)}$$

As one can see, with increasing detuning $\eta$, the denominator in front of the cosine term increases so that the amplitude of the oscillation decreases. Additionally, the frequency of the oscillations is expected to increase. Yet, we observe a constant frequency as a function of set point current for both dimers. As shown in Fig. S9a (dimer A) and b (dimer B), the frequency is obtained by fitting the data (blue curves) with a damped sinusoid (red curves). When the oscillations are not visible we fix their amplitude A to zero. The evolution of the frequency as a function of set point current is shown in Fig. S9c (dimer A) and d (dimer B) where the error bars correspond to the standard deviations of the fits.

In order to quantitatively compare the experimental results to the theoretical expectations, we should convert the change in set-point current to variations of the tip field. This is done in section S2 using the ESR measurements. However, as discussed before in Section S8, due to the temperature difference between both measurements, this can't be done fully accurately because of the influence of temperature on the spin polarization of the tunneling current. Yet, as shown in Fig. S9, a rescaling of this coefficient does not suffice to explain why the frequency remains constant: we plot in black the expected behavior for a=58 MHz/pA as determined in section S2 and in a black dashed line the expected change for a=29 MHz/pA (a very underestimated lower boundary: a factor 1/2 of the parameter.)

At the end of Section S5, we discussed several assumptions made in our model describing the coupled TiH spins. However, we do not think that these approximations can account for the absence of a frequency shift as function of detuning. As mentioned in the main text, we suspect that the constant frequency we observe could be explained using an even more rigorous description of the coupled TiH complex, taking into account the molecular bond between the Ti and H and the influence of the crystal lattice.

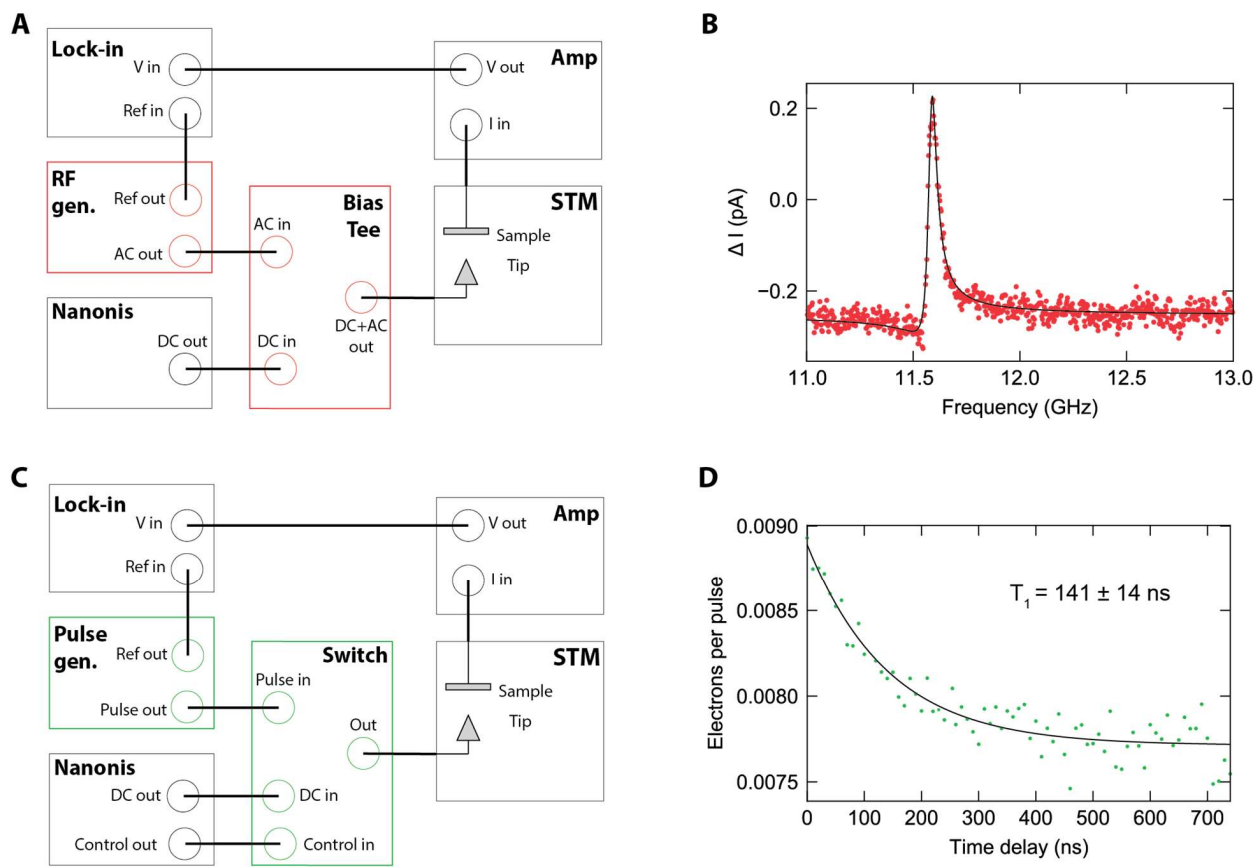

**Fig. S1.**
**Measurement setup.** (**A**) Wiring schematic for ESR measurements. (**B**) ESR resonance on a single bridge-site TiH ($T$ = 1.5 K, $V_{RF}$ = 50 mV, setpoint: 10 pA, 60 mV). Black line is a fit to a Fano function. (**C**) Wiring schematic for pump-probe measurements. (**D**) pump-probe measurement on a single bridge-site TiH ($T$ = 400 mK, $V_{pump}$ = +100 mV, $V_{probe}$ = −100 mV, pulse width 50ns, setpoint: 5 pA, 60 mV). Black line is a fit to an exponential function.

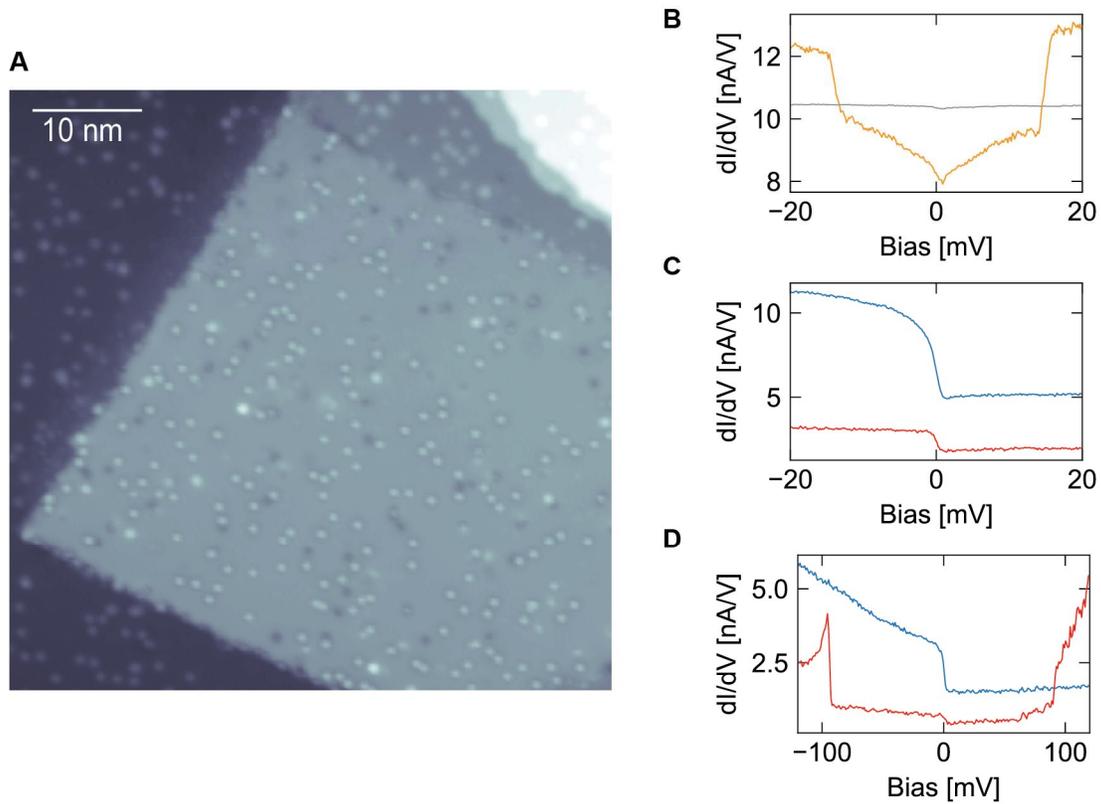

**Fig. S2.**
**Sample characterization.** (**A**) STM Topography of single Fe and TiH adatoms on a bilayer MgO island on Ag(100) (setpoint: 300 mV, 10 pA). (**B**) Conductance measurements on Fe (yellow, setpoint: 20 mV, 200 pA) and bare MgO (grey, setpoint: 20 mV, 200 pA). (**C**) Short range conductance spectra on a bridge-site TiH (blue, setpoint: 20 mV, 200 pA) and an O-site TiH (red, setpoint: 20 mV, 60 pA). (**D**) Long range conductance spectra on a bridge-site TiH (blue, setpoint: 120 mV, 500 pA) and an oxygen-site TiH (red, setpoint: 120 mV, 150 pA).

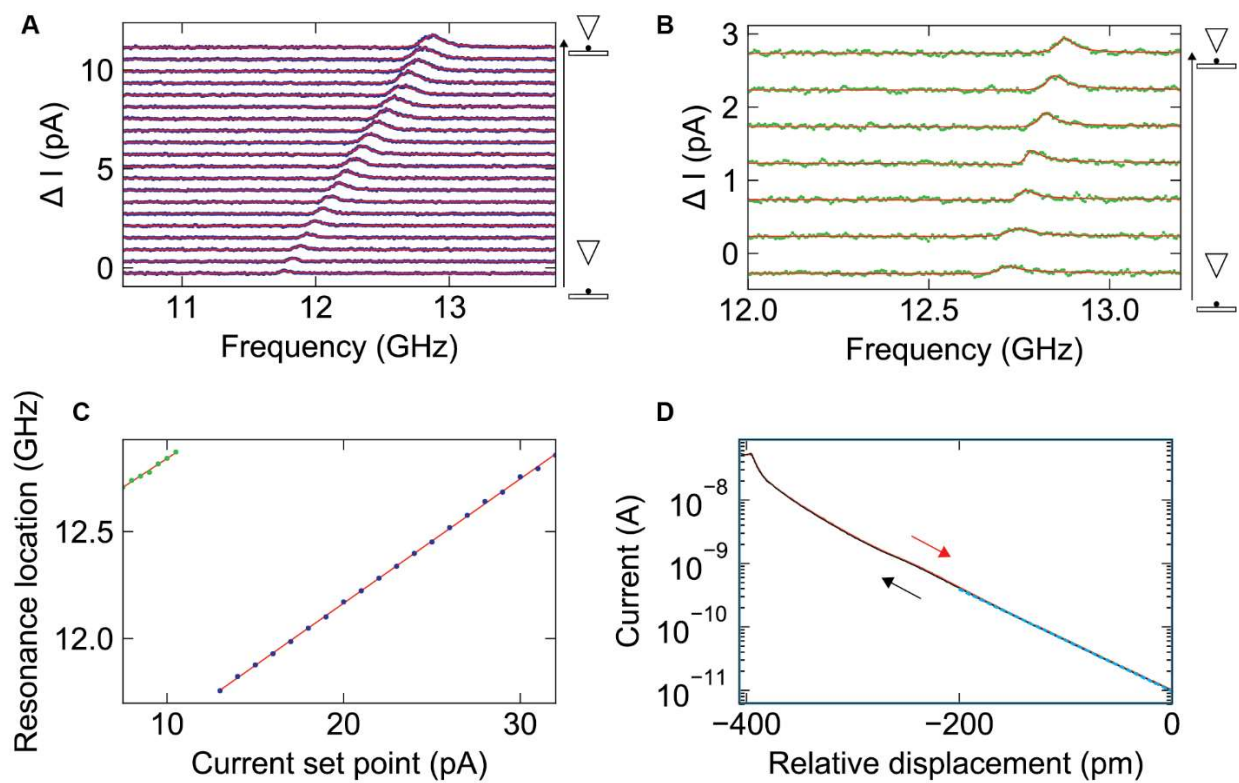

**Fig. S3.**

**Extraction of the g-factor.** (**A**) ESR spectra taken on a single TiH$_v$ at a range of setpoints: 60 mV, 13–32 pA. Red lines are Fano fits to determine the resonance location. (**B**) ESR spectra on a single TiH$_h$ (60 mV, 7.5–10.5 pA). (**C**) ESR resonances as function of setpoint for TiH$_v$ (blue) and TiH$_h$ (green). Red lines are linear fits. (**D**) Current measurement as function of tip height on top of a single bridge-site TiH (setpoint: 60 mV, 10 pA). Blue dotted line is a linear fit.

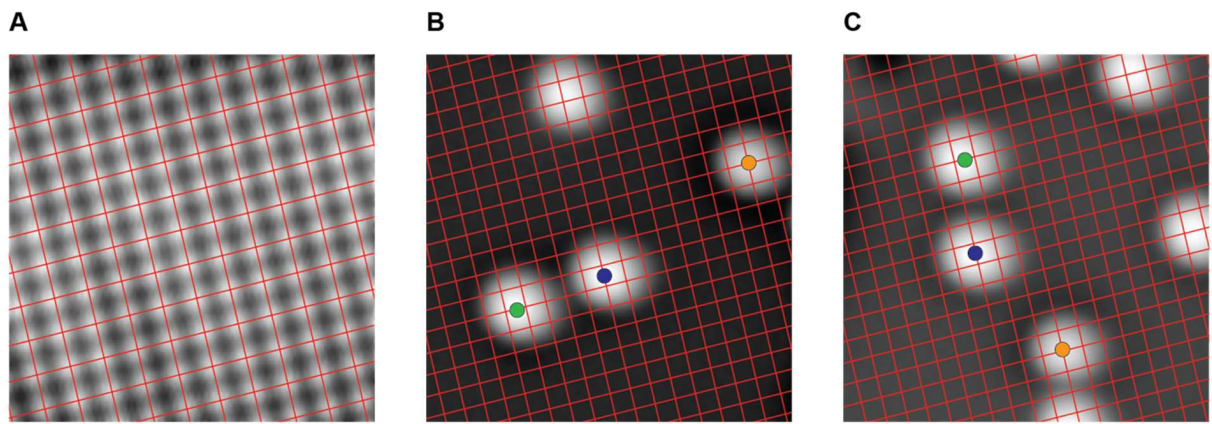

**Fig. S4.**
**Dimer placement.** (**A**) Atomic resolution topography of the MgO lattice (3 nm × 3 nm, setpoint: 5 mV, 8 nA). (**B**) Dimer B with MgO lattice overlaid (5 nm × 5 nm, setpoint: 150 mV, 10 pA). Orange dot shows an Fe atom adsorbed on an O-site. (**C**) Dimer A with MgO lattice overlaid (5 nm × 5 nm, setpoint: 150 mV, 10 pA). Orange dot shows an Fe atom adsorbed on an O-site.

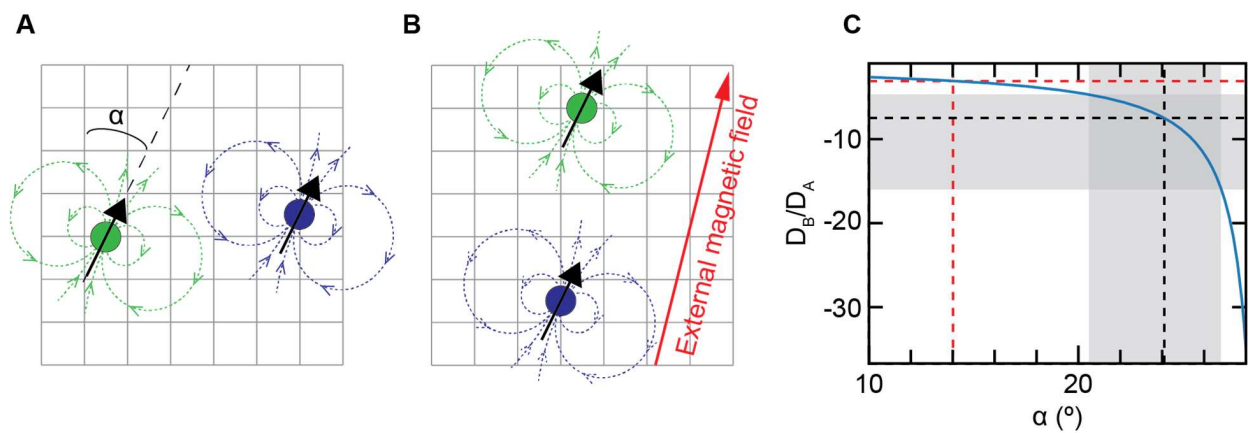

**Fig. S5.**

**Dipolar interaction.** (**A**) and (**B**): Sketches of Dimers A (B) drawn on a MgO lattice grid with the dipolar field emanating from each spin. The angle $\alpha$ indicates the angle between the lattice and the ground-state orientation of the spins, which here, in contrast to the main text, is not fixed in the direction the external field (red arrow). (**C**) Ratio of the dipolar coupling strengths of both dimers as function of $\alpha$. Black dashed lines show the ratio we find from our experiments, red dashed lines show the ratio at a 14° angle as we expect from our setup. Grey colored area shows estimated error margin.

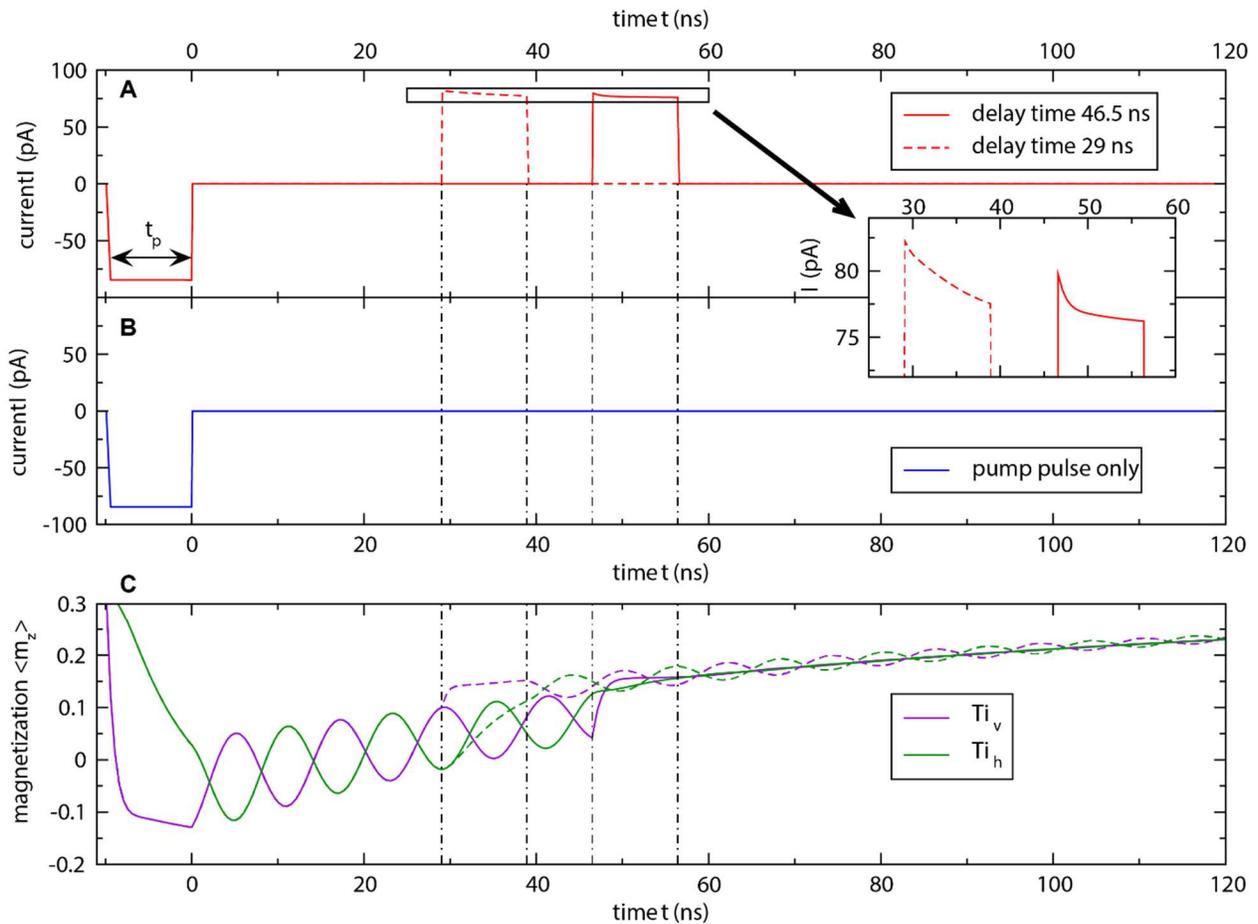

**Fig. S6.**
**Simulation of lock-in detection.** (**A**) Simulated $I(t)$ trace during a pump pulse of $V = -95.3$ mV between $t = -9.9$ ns and $0$ ns followed by a probe pulse with opposite bias sign after a delay time of 29 ns (dashed line) and 46.5 ns (full line). (**B**) $I(t)$ trace during a pump pulse alone. The difference between the traces (**A**) and (**B**) lead to the number of electrons per pulse sequence which oscillates with the delay time. (**C**) Calculated expectation value $\langle m_z \rangle$ for the $Ti_v$ and $Ti_h$ spins for the two different delay times. Inset: Zoom-in to panel (**A**).

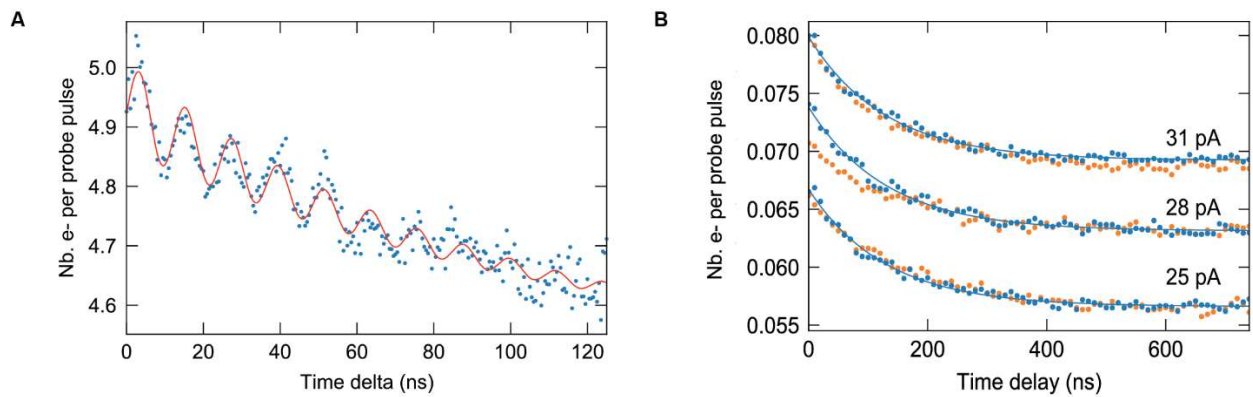

**Fig. S7.**

**Lifetime measurements on dimer A.** (**A**) Same data as plotted in Fig 3A of the main text, here with a damped sinusoidal fit to determine the decoherence time. (**B**) Pump-probe data on $TiH_h$ (blue) and $TiH_v$ (orange) of Dimer A (setpoint voltage: 60 mV, pulse width: 50 ns, pump pulse height: +100 mV, probe pulse height: −100 mV).

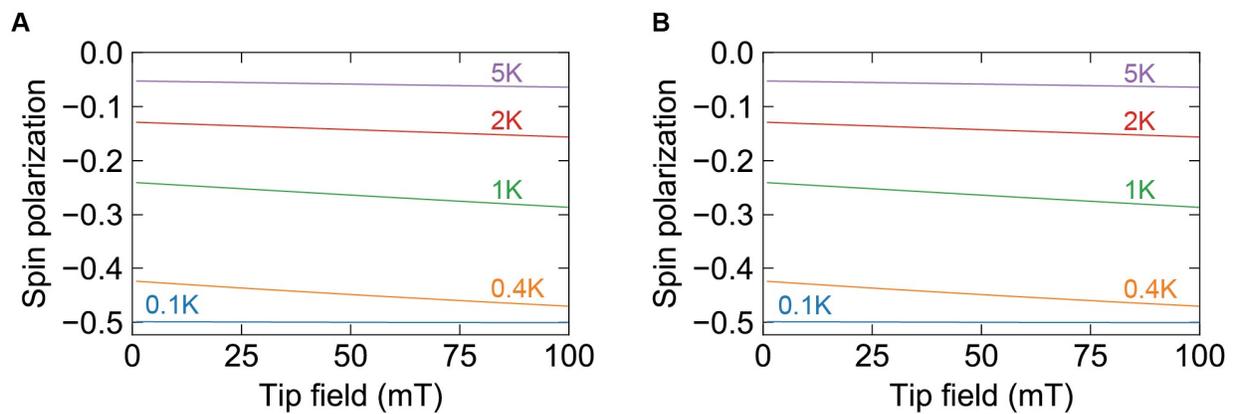

**Fig. S8.**

**Magnetization temperature dependence.** (**A**) and (**B**): Calculations on the magnetization of the ground state for dimer A (B) as function of temperature. The magnetization changes significantly between the two temperatures at which the experiments were performed: 400 mK and 1.5 K.

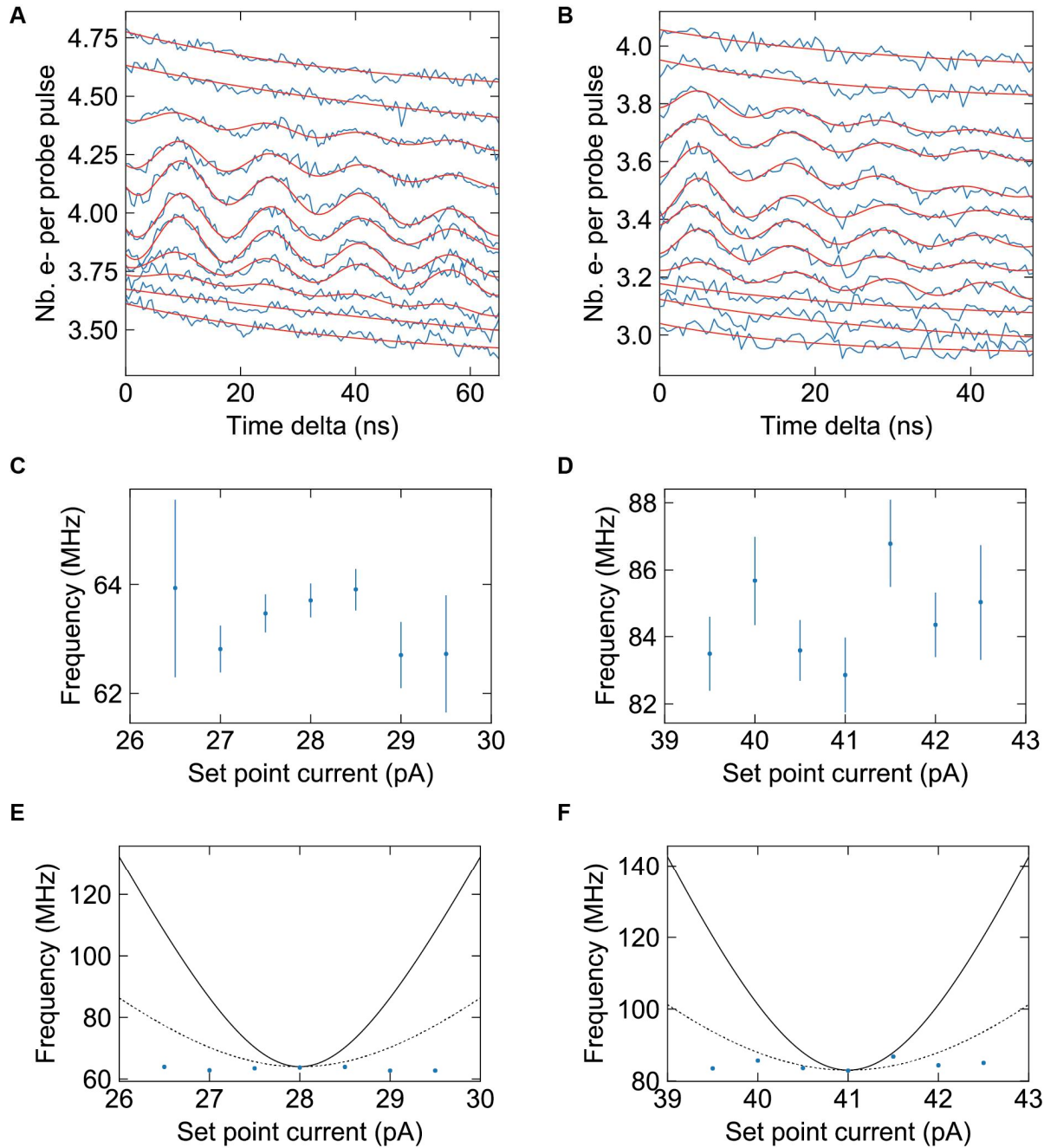

**Fig. S9.**

**Spin dynamics due to detuning.** (**A**) and (**B**): Pump probe data at different levels of detuning for dimer A (B) corresponding to the color plots show in Fig 3A and B in the main text. Setpoints: 60 mV, 25.5–30.5 pA (60mV, 38–43.5 pA), each offset by 0.05. Red lines are damped sinusoid fits where oscillations are visible and exponential fits where no oscillations are visible. (**C**) and (**D**): Extracted frequencies including error bars from the fits in (**A**) and (**B**). (**E**) and (**F**): Extracted frequencies (blue dots) plotted on top of the expected frequency shift as function of current setpoint (solid black line) and half of the expected frequency shift (dashed line) for reference.

|      | $E$ [GHz]        | $a$ [MHz/pA]   |
| ---- | ---------------- | -------------- |
| TiH$_v$ | 11.002 ± 0.005 | 58.15 ± 0.22   |
| TiH$_h$ | 12.301 ± 0.019 | 54.03 ± 2.21   |

**Table S1.**

**Fit parameters for g-factor extraction.** Fit parameters from the linear fit function (S1) to extract the g-factor of TiH$_h$ and TiH$_v$.